\documentclass[a4paper,12pt]{article}
\usepackage[pctex32]{graphics}

\begin{document}

\topmargin 0pt \oddsidemargin 0mm
\newcommand{\beq}{\begin{equation}}
\newcommand{\eeq}{\end{equation}}
\newcommand{\beqa}{\begin{eqnarray}}
\newcommand{\eeqa}{\end{eqnarray}}
\newcommand{\fr}{\frac}

\begin{titlepage}
\begin{flushright}
INJE-TP-06-03,~hep-th/0604057
\end{flushright}

\vspace{5mm}
\begin{center}
{\Large \bf Tachyon condensation and  off-shell gravity/gauge
duality} \vspace{12mm}

{\large   Yun Soo Myung \footnote{e-mail
 address: ysmyung@inje.ac.kr}}
 \\
\vspace{10mm} {\em Institute of Mathematical Science and
School of Computer Aided Science, \\Inje University, Gimhae 621-749,
Korea}

\end{center}

\vspace{5mm} \centerline{{\bf{Abstract}}}
 \vspace{5mm}
We investigate quasilocal tachyon condensation by using
gravity/gauge duality. In order to cure the IR divergence due to
a tachyon, we introduce two regularization schemes: AdS space and
a d=10 Schwarzschild black hole in a cavity. These provide stable
canonical ensembles and thus are  good candidates for the endpoint of
tachyon condensation. Introducing the Cardy-Verlinde formula, we
establish the on-shell gravity/gauge duality. We propose that the
stringy geometry resulting from the off-shell tachyon dynamics
 matches onto the off-shell AdS black
hole, where ``off-shell" means non-equilibrium configuration. The
instability induced by condensation of a tachyon behaves like an
off-shell black hole and evolves toward a large stable black
hole. The off-shell free energy and its derivative ($\beta$-function) are
 used to show the off-shell gravity/gauge duality for the
process of tachyon condensation.  Further,  d=10 Schwarzschild
black hole in a cavity is considered  for the Hagedorn transition
as a possible explanation of  the tachyon condensation.

\end{titlepage}
\newpage
\renewcommand{\thefootnote}{\arabic{footnote}}
\setcounter{footnote}{0} \setcounter{page}{2}

\section{Introduction}
In general,  closed-string tachyon condensation is a difficult
issue because we evaluate the string partition function on a cone
which has an IR divergence.  The cone is not a solution to string
equations of motion. Thus,  the Einstein equation is no longer
satisfied unless there is a delta-function source at the tip to
account for the cone curvature~\cite{Dab1}.  In this case we have
to use an off-shell formalism of string theory to compute the
partition function. In order to cure this, one considers string
theory on an orbifold ${\bf C}/{\bf Z}_N$ instead of a
cone\cite{Dab21,Dab22,Dab23,APS}. Similarly, there has been
progress  with regards to quasilocal tachyons on this direction\cite{HKMM}. For this purpose, one
introduces the winding tachyonic modes to a warped Scherk-Schwarz
cone with topology $S^1 \times Y$\cite{BB1}. A process of
winding tachyon condensation causes the circle $S^1$ on the cone
to pinch off when the size of $S^1$ reaches the string scale
$l_s$, removing the  large cone sector $X$ from the small cone
including the singularity\cite{HS}. We are interested in the
evolution of the tachyon instability in the former sector
$X$\cite{BB21,BB22}, although the latter
 can be used to study the singularity and unitarity issues in
a stringy black hole by using the tachyon condensate
phase\cite{ALMSS,MS1,MS2}.  If the dilaton can be kept small
throughout the background $X$, the process of closed-string
tachyon condensation may  be viewed either as a renormalization group (RG)-flow on the world
sheets or a flow in the space of the off-shell string backgrounds.

A black hole may be  in unstable equilibrium with a heat reservoir  in
asymptotically flat spacetime. Its fate under small fluctuations
will be either to decay to a hot flat space or to grow without
limit by absorbing radiation from the  heat reservoir. This is known as
the Gross-Perry-Yaffe (GPY) instability of a  hot flat
spacetime\cite{GPY}.  There are two ways to achieve a stable black
hole in equilibrium with heat reservoir. A black hole could be
rendered canonically stable by placing it in AdS space\cite{HP} or
by introducing a cavity surrounding it \cite{York}. This
corresponds to the IR regularization.
 The canonical ensemble for quantum
gravity in AdS$_5$ space can be defined by the Euclidean path
integral over the metric and all other fields with the time direction
periodically identified with an inverse temperature $\beta=1/T$.
In the semi-classical approach, the path integral is dominated by
configurations near saddle points, that is, classical solutions to
the Einstein equations. There are three possible
on-shell string background as saddle points: thermal AdS$_5$, a  small unstable black
hole (UBH), and a large stable black hole (SBH). The thermal AdS$_5$
of topology $S^1 \times R^4$ and SBH of topology $R^2 \times S^3$ are locally
stable, while the UBH of topology $R^2 \times S^3$ is unstable against
decay. However, all of these have a common boundary of topology
$S^1 \times S^3$ on which one may define the boundary CFT.

On the strongly coupled CFT side, string theory backgrounds of
a bubble, unstable bounce (UB)  and stable bounce (SB) appear as
saddle points in the Euclidean path integral of ${\cal N}=4, SU(N)
$ Super-Yang-Mills (SYM) theory.  According to the AdS/CFT
correspondence\cite{MGW1,MGW2,MGW3}, Witten has shown that a SBH  in
bulk-AdS$_5$ space could be described by the deconfinement phase
SB  of ${\cal N}=4$ SYM theory on $S^1\times S^3$\cite{Wit}.
Further the Hawking-Page transition corresponds to a large $N$
deconfinement/confinement transition in the strongly coupled gauge
theory. On the other hand, the UB undergoes the Gross-Witten phase
transition in the  large $N$ limit at a temperature below the
Hagedorn temperature $T=T_s$\cite{GW}. This phase transition  was
identified with the Horowitz-Polchinski correspondence
point\cite{HPo} when the size of UB becomes comparable to string
scale $(r_u \sim l_s)$\cite{AGLW}. Even for the weakly coupled
system, the phase transition is similar to  the strongly coupled
case\cite{AMMPR}.

From the microcanonical analysis, it is known that  the
d=10 Schwarzschild  black hole plays an important role at the
Hagedorn transition~\cite{BDHM,AMMPR}. The canonical instability
of the d=10 Schwarzschild black hole is well known on account of
its negative heat capacity. A gas of thermal gravitons is also
unstable to gravitational collapse due to Jeans instability. Both
processes cannot end in a large, stable black hole (sbh)
without introducing a cavity. Actually, the d=10 Hagedorn regime
is bounded by d=10 black hole and graviton phases\cite{BB1,BB21,BB22}.
For the large  't Hooft coupling $\lambda^{1/4}$, the density of
states of type IIB string theory on AdS$_5\times S^5$ has four
distinct regimes\footnote{The AdS/CFT dictionary shows
$\lambda=g_{YM}^2 N,~g_{YM}^2=4\pi g_s$, and $l=\lambda^{1/4}l_s$
with $l$ the AdS-curvature radius and radius of $S^5$.
 $S\sim E^{9/10}$ for d=10 gravitons with $E \ll \lambda^{1/4}$; $S \sim E$ for d=10
 strings with $E \gg \lambda^{5/2}$; $S \sim E^{8/7}$,  d=10
black hole with $E \gg N^2/\lambda^{7/4}$; $S \sim E^{3/4}$ for
AdS$_5$ black hole and its CFT dual with $E > N^2$\cite{AMMPR}.}.
All phases of gravitons, strings, and black holes dominated by
localized degrees of freedom in d=10 disappear, since we are
working with the canonical ensemble. However, introducing a confining
cavity with size $R$ as the IR regulator, we have a stable
canonical ensemble to study the Hagedorn transition. Then it is
possible to connect the tachyon instability to the instability of
the d=10 black hole
 that ends in a sbh.

In this work, we will describe the process of tachyon condensation
by using the off-shell gravity/gauge correspondence. For this
purpose, we use the off-shell free energy on the  bulk/boundary
sides. Since we have a hot temperature of $T=T_s$ in the decay of
the superheated Hagedorn states, the endpoint is the plasma state
of a radiated CFT (SB). On the gravity-side, this corresponds to a
SBH. Further, we use the d=10 Schwarzschild black hole in a cavity
to investigate the tachyon instability.

 The organization of this work is as follows. Section 2 is devoted to studying
 the Hawking-Page phase transition  for the nucleation of a black hole.
 The on/off-shell free energies are used to show how the phase
 transition occurs.
 The Cardy-Verlinde formula is introduced to define the ${\cal
 C}_{BH}$-function.
 In section 3,  we investigate how
 the Hagedorn transition could be used to describe  the process of quasilocal tachyon
 condensation. We use the off-shell free energy and $\beta_{BH}$-function to investigate
 the off-shell configurations.
  Inspired by the AdS/CFT correspondence, in section 4 we introduce  the Hawking-Page and Hagedorn transition
 on the CFT side  to study the process of tachyon condensation.
 In section 5, we propose  the d=10 Schwarzschild black hole in a
 cavity to describe the tachyon condensation at the  intermediate energy scale.
 Finally we summarize  our important results in section 6.

\section{Hawking-Page transition}

In this section, we review the phase transition between an AdS black
hole and thermal AdS space.    We start with the d=5
Schwarzschild-AdS black hole\cite{CTZ1,CTZ2,CTZ3,CTZ4,Wit}

\beq \label{1eq4}  ds^2_{ADSBH}=
-\Big[1+\fr{r^2}{l^2}-\fr{m}{r^2}\Big]dt^2
+\fr{dr^2}{1+\fr{r^2}{l^2}-\fr{m}{r^2}}+r^2d\Omega_{3}.
 \eeq
Here the reduced mass $m=r_+^2(1+r_+^2/l^2)$ is determined by the
horizon radius $r_+^2=(l^2/2)[-1+\sqrt{1+4m/l^2}]$.  The ADM mass
$M=E_{BH}^{A}$ is related to the reduced mass $m$ as $M=\fr{3V_3
m}{16 \pi G_5}$ with $V_3=2\pi^2$ the volume of unit three sphere.
It possesses a continuous mass spectrum from $M(l,r_+)$ to thermal
AdS with $M(l,0)=0$: \beq \label{ADS}
ds^2_{ADS}=-(1+r^2/l^2)dt^2+(1+r^2/l^2)^{-1}dr^2+r^2d\Omega_{3}^2.
\eeq Its Hawking temperature and heat capacity are given by\cite{MYU} \beq
\label{2eq4} T^{A}_H(l,r_+)=\fr{1}{2\pi }
\Big[\fr{1}{r_+}+\fr{2r_+}{l^2}\Big],~~C_{BH}^{A}(l,r_+)=\frac{3V_3r_+^{2}}
{2G_5}\Big[\fr{r_+^2+r_0^2}{r_+^2-r_0^2}\Big]\eeq with
$r_0=l/\sqrt{2}$.  $T^{A}_H$  has the minimum value of
\cite{HP}\beq \label{3eq4} T_0=\frac{\sqrt{2}}{\pi l} \eeq at the
minimum length $r_+=r_0$, and grows linearly for large $r_+$ due
to the presence of a negative cosmological constant
$\Lambda=-6/l^2$. A solution to the thermal equilibrium (on-shell)
condition of $T_{H}^{A}=T$ corresponds to small, unstable black
hole (UBH) with radius $r_u=(\pi l^2T/2)[1-\sqrt{1-8/(2\pi
lT)^2}] \le r_0$, while one is large, stable black hole (SBH) with radius
$r_s=(\pi l^2T/2)[1+\sqrt{1-8/(2\pi lT)^2}] \ge r_0$. Here we find an
inequality of the temperature $T\ge T_0$ and  a sequence of
$r_u<r_0=35<r_s$ for $l=50$. For $T \gg 1/l$, we have approximate
on-shell relations  of $r_u \simeq 1/2\pi T$ and $r_s \simeq \pi
l^2 T$. In the  off-shell case  of $T\not=T^A_H$, there is no
direct relation between $r_+$ and $T$. In this sense, we may
regard $r_+$ as the effective temperature. For $T<T_0$, there is
no solution to $T_{H}^{A}=T$, which means that no black hole can
exist in AdS space. The heat capacity in Eq.(\ref{2eq4}) has an
unbounded discontinuity at $r_+=r_0$, signaling a first-order
phase transition from negative heat capacity to positive one.
However, the heat capacity determines the thermal stability of a
system: thermally stable (unstable) if
$C_{BH}^{A}>0(C_{BH}^{A}<0)$.
 We get an important piece of  information
from the study of the BH-free energy \beq \label{4eq4}
F^{on}_{BH}(l,r_+)=E_{BH}^{A}-T^{A}_H S_{BH}^{A} \equiv
F^{on}_{UBH}+F^{on}_{SBH},\eeq where \beq \label{5eq4}
S_{BH}^{A}=\frac{V_3}{4 G_5}
r_+^{3},~F^{on}_{UBH}=\frac{V_3r^2_+}{16 \pi
G_5},~F^{on}_{SBH}=-\hat{r}^2F^{on}_{UBH}\eeq  with
$\hat{r}=r_+/l$. This applies to saddle points of $r_u$ and $r_s$
only. We observe a change of sign between
 $F^{on}_{BH}(l,r_+ \ll l)\simeq F^{on}_{UBH} \sim r_+^2$ and $F^{on}_{BH}(l,r_+ \gg l)\simeq F^{on}_{SBH} \sim
-r_+^{4}$. This shows that
 an UBH  is unstable to decay into
thermal AdS space, while  a SBH is stable against decay. For
$r_+ \gg l$, we have an approximate entropy-energy relation of
$S_{BH}^{A}\sim (E^{A}_{BH})^{3/4}$, which means that a large
black hole has a stable canonical ensemble. Here we have the
transition point of $r_+=r_1=l$ from $F^{on}_{BH}(l,r_+)=0$.
Plugging this into Eq.(\ref{2eq4}) leads to the transition
temperature\cite{HP,CTZ1,CTZ2,CTZ3,CTZ4} \beq \label{6eq4} T_1=\frac{3}{2\pi l}
\eeq which is the critical temperature for the Hawking-Page phase
transition.

Assuming the CFT-dual, we introduce the Casimir energy for the
black hole\footnote{Frankly, it is hard to define the bulk-Casimir energy
unless assuming the presence of its dual CFT.
Hence it is fairly asserted that the Cardy-Verlinde formula in Eq. (\ref{BCV}) on the bulk-side comes from
its Cardy-Verlinde formula in Eq. (\ref{6eq7}) on the boundary.}~\cite{KPSZ} \beq \label{BCE} E_{BH}^{c}\equiv
4E_{BH}^{A}-3T_{H}^{A}S_{BH}^{A}=\frac{3V_3r_+^2}{8\pi G_5}. \eeq
It allows us  to write the Cardy-Verlinde formula on the
bulk-side\cite{Ver} \beq \label{BCV} S_{BH}^{A}= \frac{2 \pi
l}{3}\sqrt{E_{BH}^c\Big(2E_{BH}^{A}-E_{BH}^c\Big)}. \eeq This is
an exact relation between  entropy and energy. Minimization with
respect to $E_{BH}^c$ leads to the bound $S_{BH}^{A} \le
(1/T_1)E_{BH}^{A}$, which is called  the AdS-Bekenstein bound. This
is not only the upper bound for $S_{BH}^{A}$, but the lower bound
for $E_{BH}^{A}$. The saturation of this bound is achieved at the
transition point $r_+=r_1$. Here we have  thermal
relations:
\begin{eqnarray} \label{BCV1}
 &E_{BH}^{A}&= E^c_{BH}=S_{BH}^{A}=0,~F^{on}_{BH}>0, ~~{\rm
 for}~
 r_+<r_1\\ \label{BCV2}
&E_{BH}^{A}&=E^c_{BH}=T_1S_{BH}^{A},~F^{on}_{BH}=0,~~{\rm
for}~r_+=r_1 \\  \label{BCV3}
  &E_{BH}^{A}& >E^c_{BH},~
S_{BH}^{A}<(1/T_1)E_{BH}^{A},~F^{on}_{BH}<0, ~~{\rm for}~ r_+>r_1.
\end{eqnarray} This means that the Cardy-Verlinde formula is valid
only  for on-shell with $r_+ \ge r_1$, that is, for a SBH.

Further, defining the Casimir entropy as
$S_{BH}^c=E_{BH}^c/T_1=V_3lr_+^2/4G_5$, the BH-free energy takes
the form \beq \label{onf} l
F^{on}_{BH}=\frac{S_{BH}^c}{4\pi}\Big[1-\hat{r}^2\Big]. \eeq
 If $\hat{r} \to  \delta^{-1}$, this is
analogous  to the d=2 free energy of
$lF_2=\frac{c}{24}\Big(1-\delta^{-2}\Big)$ for $c$ free bosons.
Here we find a relation  $c= 6S_2/\pi$ between the central charge
$c$ and the Casimir entropy $S_2$ in CFT$_2$. In this work we
choose a slightly different notation to go together  with its dual
CFT. The ${\cal C}_{BH}$-function is defined as \beq \label{cfunc}
{\cal C}_{BH}(l,r_+)\equiv \frac{S_{BH}^c}{4\pi}=\frac{V_3 l
r_+^2}{16\pi G_5},~r_+ \ge r_1\eeq which is useful for describing
the bulk RG-flow in the Hagedorn transition. From Eq.(\ref{BCV1}),
one finds ${\cal C}_{BH}(l,r_+)=0$, for $0 \le r_+ <r_1$.

In order to discuss the phase transition more explicitly, we
introduce the generalized free energy $F^{off}_{BH}= E_{BH}^{A}-
T S_{BH}^{A}$, which applies to any value of $r_+$ with the
fixed $T$\cite{York,BB21,BB22}. It is given by\beq \label{8eq4}
F^{off}_{BH}(l,r_+,T)=F^{off}_{UBH}+F^{off}_{SBH}, \eeq where\beq
\label{schf}
F^{off}_{UBH}=3F^{on}_{UBH}\Big[1-\frac{2}{3}\frac{T}{T^{A}_H}\Big],~~
F^{off}_{SBH}=-3F^{on}_{SBH}\Big[1-\frac{4}{3}\frac{T}{T^{A}_H}\Big]
\eeq Its connection to the action  is given by $I_{BH}=\beta
F^{off}_{BH}$. In case of $r_+\ll l ~( \hat{r} \ll 1)$, we
approximate $F^{off}_{BH}$ by $F^{off}_{UBH}$, while for $r_+\gg
l$ $(\hat{r} \gg 1)$, it is approximated to be $F^{off}_{SBH}$. As
is shown in Fig. 1, for $T=T_0$, an extremum appears at
$r_+=r_0(=r_u=r_s)$. We confirm that for $T>T_0$, there are two
saddle points, UBH with radius $r_u$ and SBH with radius $r_s$.
$F^{on}_{BH}$ is composed of  a set of two saddle points for
$F^{off}_{BH}$. That is, $F^{on}_{BH}$ can be obtained from
$F^{off}_{BH}$ through the operation: $\partial
F^{off}_{BH}/\partial r_+=0 \to T=T^{A}_H \to
F^{off}_{BH}=F^{off}_{BH}$. Hence the BH-free energy $F^{on}_{BH}$
is regarded as the on-shell (equilibrium) free energy, whereas the
generalized free energy $F^{off}_{BH}$ corresponds to the
off-shell (non-equilibrium) free energy. All saddle points
including thermal AdS  at $r_+=0$ contribute dominantly to the
path integral evaluation for canonical ensemble.

\begin{figure}[t!]
 \centering
   \includegraphics{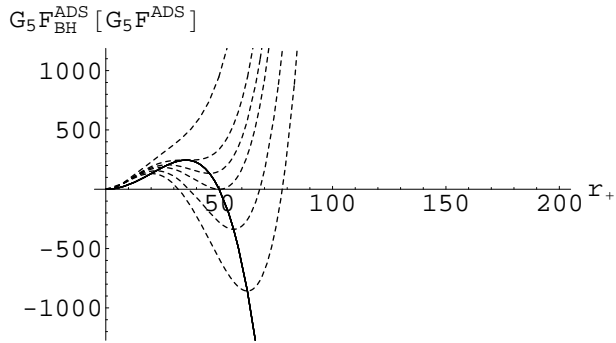}
    \caption{The process of   black hole nucleation as in the Hawking-Page phase transition.
     Here $r_+$ plays a role of the effective
    temperature for the BH-free energy, but it is merely
    a coordinate for the generalized free energy with the fixed
    $T$.
   The solid line represents the BH-free
energy $F^{on}_{BH}(l=50,r_+)$ in the units of $G_5$,  while the
dashed line denotes the generalized free energy
$F^{off}_{BH}(l=50,r_+,T)$ for six different temperatures: from
the top to bottom, $T=0.008, T_0(=0.009), 0.0093, T_1(=0.0095),
0.01,0.0105$.} \label{fig1}
\end{figure}

At this stage, we briefly describe the Hawking-Page
transition\cite{HP}. For $T_0<T<T_1$, we have a sequence (see the
third dashed graph in Fig. 1) \beq \label{9eq4}
F^{off}_{BH}(r_+=0)=0<F^{off}_{BH}(r_+=r_s)
<F^{off}_{BH}(r_+=r_u)\eeq
 which means that the saddle point
for  thermal AdS dominates. For $T>T_1$, the SBH dominates because
of $F^{off}_{BH}(r_+=r_s)<0$. There is a change of dominance at
the critical temperature $T=T_1$. The UBH plays a role of the
mediator for the transition between thermal AdS and SBH. In case
of $T<T_1$, the system is described by a thermal gas, whereas
above $T_1$ it is described by a SBH.  This is the Hawking-Page
transition for black hole nucleation.

Finally, we introduce the classical gravitational effect of
thermal radiation in
 AdS space. From a relation of $T^5_2r_+^4 \sim r_+^4/G_5l^2$ for  $r_+ \gg l$,
 we find  the
 collapsing temperature $ T_2=1/(G_5l^2)^{1/5}$.
 Thermal radiation at $T>T_2$ would not be
able to support itself against its self-gravity and thus it would
collapse to form a black hole. The collapsing temperature of
$T_2>T_s$ is much  greater than $T_0$ and $T_1$ because of
$(G_5l^2)^{1/5}\sim l/N^{2/5}< 1$. Hence, in this work, we exclude
the gravitational collapse of thermal AdS from our consideration.

\section{Hegedorn transition and quasilocal tachyon condensation}

We  start with  the Hagedorn   transition in the type IIB
string theory on AdS$_5 \times S^5$.  A good model for the
Hagedorn spectrum is to consider the highly excited strings as a
random walk. Its microcanonical density of states
 takes the form
 \beq \label{density}
 \Omega(E)={\rm exp}(\beta_s E +\cdots), \eeq
  where the ellipsis
 is $c\sqrt{E}(-c^\prime E^{16}e^{-\eta E})$ for open strings (closed strings).

 According to the AdS/CFT correspondence, Witten has shown
that a SBH in AdS space could be described by the deconfinement
phase of ${\cal N}=4$ SYM theory on $S^1\times S^3$. Further, the
Hawking-Page transition corresponds to a large $N$ deconfinement
transition in the strongly  coupled gauge theory. At strong
coupling, the Hagedorn temperature $T=T_s=1$ is much greater than
the critical temperature $T_1 =3/2 \pi l$ for the Hawking-Page
transition.
 According to  Hagedorn censorship, it was proposed
that the regime of Hagedorn transition cannot be studied in terms
of the Hawking-Page transition. This is because although the
Hagedorn phase is stable microcanonically, but it is unstable
canonically. Once the temperature reaches  the Hagedorn
temperature, the hotter heat reservoir pumps energy into the
system until it reaches a large, stable AdS black hole.

 However,
we use this property to describe the Hagedorn tachyon
condensation. For this purpose, we use the winding  tachyonic
modes from a warped Scherk-Schwarz compactification with topology
$S^1 \times Y$.  The winding tachyon instability causes
the circle $S^1$ on the cone to pinch off when the size of $S^1$
reaches the string scale $l_s$, removing the cone of large
curvature sector $X$  from the small cone including the
singularity. We are interested in the evolution of the tachyon
instability in the former sector $X$, although the latter
 can be used to study the singularity and unitarity issues in
the stringy black hole by using the tachyon condensate phase. The
former process is related  to the formation and growth of an
off-shell black hole with its near horizon geometry ($<$), while
the latter process is considered as the formation and evaporation
of an off-shell black hole with its shape ($<>$) on the gravity
side. If the singularity is replaced by the tachyon phase\cite{MS1,MS2},
it shape is given by $|T>$.  If the dilaton could be kept small
throughout the background $X$, the process of the closed-string
tachyon condensation can be viewed either some RG-flow on the
world sheet or off-shell bulk/boundary flows. Let us introduce the
AdS regularization for a string gas in AdS$_5 \times S^5$ of type
IIB string theory, whose metric is given by \beq \label{1eq5}
ds^2_{10ADS}= ds^2_{ADS}+ l^2 d\Omega_5^2
 \eeq
where $l=\lambda^{1/4}$ \footnote{ The d=10 Newton constant is
given by $G_{10}=2^3\pi^6 g_s^2l_s^8$ and its relation to the d=5
Newton constant is $G_{10}=V(S^5)G_5=\pi^3 l^5G_5$. Then
$N^2=l^8/(16\pi^2g_s^2l_s^8)=\pi l^3/2G_5$\cite{Gub1,Gub2}. For a
numerical computation, we use string units: $l_s=
\sqrt{\alpha^\prime}=1$. A strongly coupled regime is chosen to be
$g_s =\pi> 1$. For $l=\lambda^{1/4}=50$, $N^2=2.5\times
10^{10},~G_{10}=75908,~ G_5=7.8 \times
10^{-6},~(G_5l^2)^{1/5}=0.455<1$. The radius of Jeans instability
at string scale is $r^{J}_{st}=l_s/g_s=0.318$.}with $N$ the
quantum number of Ramond-Ramond flux on $S^5$. The time circle in thermal
AdS$_5$ is non-contractible and thus the winding number is
preserved.
   Hence we propose that the stringy geometry resulting
from the off-shell tachyon dynamics with $\sigma \sim m_sr_c$
matches onto the off-shell AdS black hole with the horizon radius
$r_+ \sim r_{st}\in \{l_s,l\}$. Here $\sigma$ is an effective tachyon, $m_s=1/l_s$ is the mass of string,
and $r_c$ is  a capping radius. $X_{st}$ is introduced to denote the corresponding
Euclidean geometry with a conical singularity at the event
horizon. Hence its near horizon geometry takes a shape of ``$<$".
Then the process of the off-shell black hole growth provides a good
way to escape  from the Hagedorn regime of $0\le r_+ \le l$. That
is, the off-shell black hole continues to grow until it arrives at
a SBH with $r_+ = r_s$. The
 endpoint geometry $X_{s}$ is the Euclidean section of the
AdS black hole with metric  \beq \label{2eq5} ds^2_{10ADSBH}=
ds^2_{ADSBH}+ l^2 d\Omega_5^2.
 \eeq
The on-shell manifold $X_{s}$  has topology of $R^2 \times S^3
\times S^5$, which is considered as a capping of $X_c$ at
$r_+=r_s$\cite{BB1,BB21,BB22}. Here $X_c$ corresponds to the Euclidean
off-shell black hole with a conical singularity ($<$). Hence  its near horizon geometry takes a
shape of ``$\subset$", where there is no singularity at the event
horizon. The capping is a progressive effect if $X_c \in
(X_{st},X_{s})$.  This is a
topology changing process, where we have non-equilibrium
configuration  $T \not= T^{A}_H$.

 At this stage, we
introduce the off-shell parameter $\alpha$ related to the deficit
angle $\delta$\cite{FFZ}
 \beq \label{off-shell} \alpha(l,r_+,T) =
\frac{T_{H}^{A}(l,r_+)}{T} \equiv 1-\frac{\delta(l,r_+,T)}{2\pi}.
\eeq  Then the off-shell free energy takes the form \beq
\label{defree}
F^{off}_{BH}(l,r_+,T)=\frac{F^{on}_{BH}}{\alpha}+\frac{(\alpha-1)}{\alpha}E_{BH}^{A},
\eeq where the first term is similar to the on-shell free energy,
while the second one determines the off-shell nature of the free
energy. Here the corresponding action is given by $I_{BH}=\beta F^{off}_{BH}$.
For $\alpha=1(T=T^{A}_H,~\delta=0)$, we recover two saddle
points which satisfy $F^{off}_{BH}=F^{on}_{BH}$. Hence, for fixed
$T\not=T^{A}_H$, $F^{off}_{BH}$ describes off-shell configurations
very well.  As is shown in Fig. 2, the off-shell parameter
$\alpha$ behaves like the temperature of a cool (off-shell) black
hole but the deficit angle $\delta$
 has the maximum at $r_+=r_0$ and deceases to zero at $r_+=r_s$.  $\delta(X)$
classifies a conical singularity at the event horizon. For
example, $\delta(X)=0$ for on-shell configurations of
$X=X_u,~X_{s}$ without any cone ($\subset$), while it has the maximum value
$\delta(X)=2 \pi(1-T_0)\simeq 2\pi$ for $X=X_0$ with the narrowest
cone ($\prec$). In general, we have $0<\delta(X)<2\pi$, for the geometry  $X \in
(X_u,X_s)$ with a cone $(<)$.

\begin{figure}[t!]
 \centering
   \includegraphics{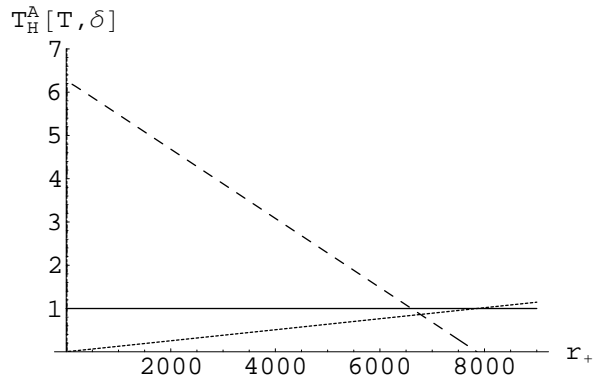} \caption{The temperature picture of a cool (off-shell) black hole growth
  in a hotter heat bath at $T>T^A_{H}$.  Solid
line: hotter temperature of heat reservoir at Hagedorn temperature
$T=T_s=1$.
  Dotted line: plot of the  increasing temperature $T^{A}_{H}$ of a cool
   black hole with $l=50$. Also, this graph indicates the off-shell parameter
   $\alpha(l=50,r_+,1)$. Dashed line denotes the deficit angle $\delta(l=50,r_+,1)$.
   In this case we have
the sequence of points : $r_u=0.159<r_0=35<r_s=7853$. If a
matching
  occurs at $r_+=r_{st} \ge l_s$, the stringy black hole always grows into a SBH at $r_+=r_s$.} \label{fig2}
\end{figure}

Our analysis is  based on the heat capacity of the black hole.
Neglecting quantum tunneling process\cite{CTZ1,CTZ2,CTZ3,CTZ4}, we have two cases
only. i) If the initial black hole state satisfies $r_+<r_u$,
there is no black hole state because the   black hole evaporates
until it arrives at its final state of thermal AdS. ii) If
$r_+>r_u$ initially, this black hole grows into  a SBH with size
$r_+=r_s$. Hence, if the stringy black hole is formed with  a size
$r_{st}\ge l_s$, this
 grows into a globally stable black hole  with $r_{s} \simeq \pi l^2
T_s=7853$ by absorbing the remaining string gas in AdS space. This
feature is depicted  in Fig. 2.

\begin{figure}[t!]
 \centering
   \includegraphics{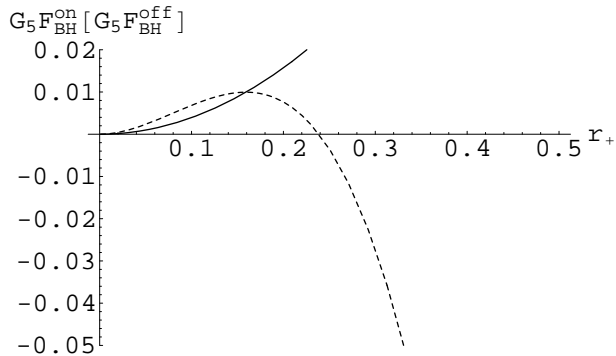} \caption{Hagedorn transition with $T=T_s$ for $r_+ \ll l$.
   The solid line represents the on-shell free
energy $F^{on}_{BH}(50,r_+)\simeq F^{on}_{UBH}(r_+)$. The dashed
line denotes the off-shell free energy
$F^{off}_{BH}(50,r_+,1)\simeq F^{off}_{UBH}(r_+,1)$ in the units
of $G_{5}$. This picture is described by an UBH. A
junction point located at $r_+=r_u=0.159$ is the maximum. }
\label{fig3}
\end{figure}
\begin{figure}[t!]
 \centering
   \includegraphics{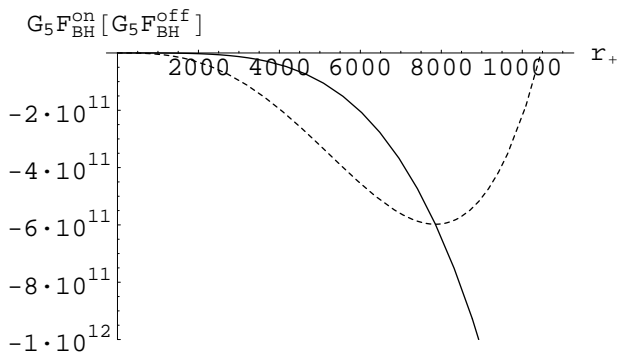} \caption{The free energy picture of a cool (off-shell) black hole growth
  in a hotter heat bath for $r_+ \gg l=50$.
   The solid line represents the on-shell free
energy $F^{on}_{BH}(50,r_+)\simeq F^{on}_{SBH}(50,r_+)\sim
-r_+^4$, while the dashed lines denote the off-shell free energy
$F^{off}_{BH}(50,r_+,1)\simeq F^{off}_{SBH}(50,r_+,1) \sim r_+^4$
in the units of $G_{5}$. A junction point is located at
$r_+=r_s=7853$.} \label{fig4}
\end{figure}
\begin{figure}[t!]
 \centering
   \includegraphics{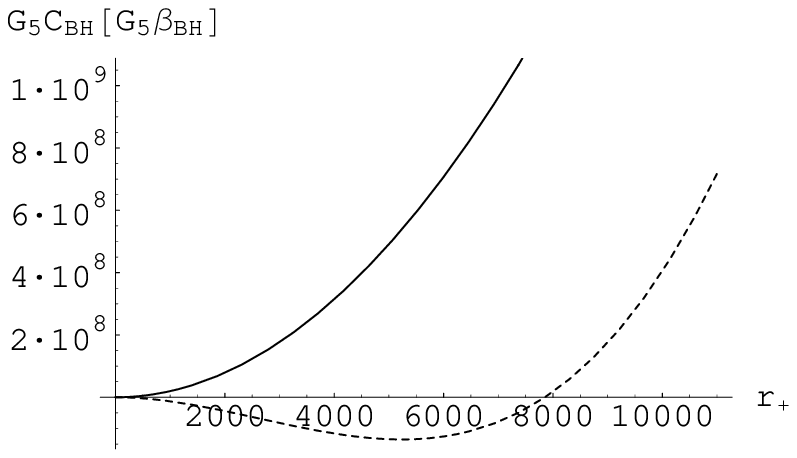} \caption{The ${\cal C}_{BH}$ versus $\beta_{BH}$-function.
   The solid line represents the on-shell ${\cal C}_{BH}(50,r_+\ge r_1)$-function $\sim r_+^2$,
while the dashed line denotes the off-shell $\beta_{BH}(50,r_+,1)$
in the units of $G_{5}$. The $\beta_{BH}$-function is just the derivative of $F^{off}_{BH}$
with respect to $r_+$. Thus it is zero at the saddle points of
$r_+=r_u,r_s$. } \label{fig5}
\end{figure}

In our picture, the generalized free energy in Eq. (\ref{defree})
can be used to explain this non-equilibrium process.  Here we
choose the free energy for thermal AdS $(X_{ADS})$ to be zero.
Considering $X_c\in \{X_{st},X_s\}$, $X_c$  has the conical
singularity and its off-shell free energy is given by a point
$F^{off}_{BH}(X_c)$ on the dashed graph in Fig. 3 for  $r_+ \ll l$
and Fig. 4 for $r_+\gg l$. As the initial off-shell black hole
($F^{off}_{BH}(X_{st})$)
 flows into the final  black hole
($F^{off}_{BH}(X_s)$),  an off-shell free energy curve
($F^{off}_{BH}(X_c)$) connects $X_{st}$ to $X_{bh}$.

 Barbon and Rabinovici\cite{BB1} assumed that, for  $T \ge T_s$, the
barrier of the UBH is completely washed out by the tree-level
string effect. In this case, at $r_+=0$, one sees the
tree-level instability of the Hagedorn tachyon. From Fig. 3, we
observe the UBH barrier
 at $r_u=0.159$, even though its maximum is small as
$G_{5}F^{off}_{UBH}(0.159,1)=0.009$. Also we observe the
sensitivity around $r_+=l_s=1$ that
$G_{5}F^{off}_{UBH}(0.238,1)=0,~G_5F^{off}_{UBH}(r_{st}^J=0.318,1)=-0.039,~G_5F^{off}_{UBH}(1,1)=-3.75,$
and $G_5 F^{off}_{UBH}(2,1)=-34.75$. The distance between the
maximum and zero is very small as $\Delta r_+=0.079$. This shows
that the UBH barrier suffers from ${\cal O}(1)$ ambiguity at the
Hagedorn temperature. Hence we may  take a matching point at
$r_{st}\gg l_s$. From the relation of $r_u \simeq 1/2\pi T$, we
conjecture that thermal AdS and UBH will merge for
large $T$.

Inspired by the action relation of $I_{BH}(l,r_+=r_c,T=T_s) \simeq
\beta l^9 V_{eff}^{t}(|\sigma|=m_s r_c)$ with $r_c$ the capping
radius and $m_s=1/l_s$ the string mass\cite{BB1,BB21,BB22},  the effective tachyon
potential $V_{eff}^{t}$ is given by \beq \label{7eq7}
V_{eff}^{t}(|\sigma|=r_c)\simeq
\frac{F^{off}_{BH}(l=50,r_+=r_c,T=1)}{l^9}. \eeq The
corresponding graph of $V_{eff}^{t}$ appears in Fig. 3 for
small $r_+ \ll l$. For  large $r_+ \gg 1$, see Fig. 4.
This is an approximation to the Atick-Witten effective potential for the
thermal tachyon on $S^1\times R^9$\cite{AW}.

 As is shown in
Fig. 4, at $r_+=r_s=7853$, the endpoint of Hagedorn
tachyon condensation is  a globally stable black hole state in AdS
space.  The monotonic property of ${\cal C}_{BH}$
provides  condition  \beq \label{cfunb} T_{H}^{A} \frac{d{\cal
C}_{BH}(50,r_+)}{dT_{H}^{A}} \ge 0 \to r_+ \ge r_0, \eeq which
coincides with the positive heat capacity of $C_{BH}^{A} \ge 0$.
Combining it with Eq. (\ref{cfunc}) leads to the monotonicity:
$\Delta {\cal C}_{BH}(l,r_+\ge r_1)\ge 0$.  This shows that there
exists a strong relationship between the monotonicity of ${\cal C}_{BH}$ and the thermal stability of a SBH. In general,
the ${\cal C}_{BH}$ is closely related to the central
charge on the boundary CFT. From Fig. 4 and 5,  we confirm the
on-shell monotonicity connection between the increasing ${\cal
C}_{BH}(l,r_+)$  ($\nearrow$) and the decreasing $F^{on}(l,r_+)$ ($\searrow$).

Next, we study the off-shell feature.  At  fixed
temperature we move along the renormalization group trajectories.
From the definition of the bulk $\beta$-function, we have \beq
\label{betafucn} \beta_{BH}(l,r_+,T) \propto \frac{\partial
I_{BH}}{\partial r_+}=-\frac{6{\cal
C}_{BH}(l,r_+)}{l}\delta(l,r_+,T). \eeq As is shown in Fig. 5,
the bulk $\beta$-function does not show a monotonic feature and  measures
the mass of the conical singularity at the event horizon
\footnote{Considering $I_{BH}=F^{on}_{BH}/T^A_H+I_{cs}$
, we can define the mass $M_{cs}$ of a conical singularity as $I_{cs}=4\pi r_+ M_{cs}$\cite{KPR,myung3}.
Then it is given by $M_{cs}=\Big[\frac{r_+^2+l^2}{2r_+^2+l^2}\Big]\frac{\beta_{BH}}{8\pi}$
with $\beta_{BH}=-\frac{3 \pi r_+^2}{4G_5} \delta$. In  case of $r_+ \gg l$, we have an approximate
relation between the $\beta$-function and the mass of the conical singularity:
$M_{cs}\simeq \frac{\beta_{BH}}{16\pi}$.}.

The off-shell free energy $F^{off}_{BH}(l,r_+,T)$ describes
the non-equilibrium configurations between $r_u$ and $r_s$. This
is a monotonically decreasing function between  on-shell
configurations of $X_u$ and $X_s$ only. Hence we confirm the
off-shell correspondence between
the decreasing  $F^{off}_{BH}(l,r_+,T)$ ($\searrow$) and the $\beta$-function in the
shape of $\searrow \nearrow$.

\section{Confining/deconfining and  Hagedorn  transitions on the strongly coupled CFT side}

In this section, we are interested in the confining/deconfining and  Hagedorn  transitions for
a strongly coupled, large $N$ gauge theories in the  high
temperature limit.  The d=4 CFT on $R \times S^3$ near infinity is
defined by the Einstein static universe~\cite{Ver}: \beq
\label{1eq7} ds_{ESU}^2=-d\tau^2+\rho^2d\Omega_3^2,\eeq where
$\tau$ is a time variable in the boundary theory and $\rho$ is the
radius of three sphere $S^3$. Now we can determine the above
boundary metric by using the bulk metric in Eq.(\ref{1eq4}) near
infinity as
 \beq
\label{2eq7} ds_b^2=\lim_{r\to
\infty}\frac{\rho^2}{r^2}ds_{ADSBH}^2 =-d\tau^2+\rho^2d\Omega_3^2
\to ds_{ESU}^2,~~ \tau=\frac{\rho}{l}t. \eeq Using the Euclidean
formalism on $S^1 \times S^3$, we find  bulk-boundary
relations~\cite{CMN}: \beq \label{3eq7} T_{CFT}=\frac{l}{\rho}
T^{A}_H=\frac{1}{2 \pi \rho}\Big[2\hat{r}+\frac{1}{\hat{r}}\Big],~
E_{CFT}=\frac{l}{\rho} E_{BH}^{A}=\frac{3V_3\kappa
\hat{r}^2}{\rho}\Big[1+\hat{r}^2\Big] \eeq and \beq \label{4eq7}
F^{on}_{CFT}=\frac{l}{\rho}F^{on}_{BH}=\frac{V_3\kappa\hat{r}^2}{\rho}\Big[1-\hat{r}^2\Big],
~S_{CFT}=4 \pi \kappa V_3 \hat{r}^3=S_{BH}\eeq with $\kappa=l^3/16
\pi G_5$.
These are a realization of the holographic principle via the AdS/CFT correspondence.
We note that all of CFT's quantities are locally measured in the dual CFT, even though the bulk quantities
are observed at infinity. Hence  the AdS-curvature radius $l$ disappears and instead
the boundary radius $\rho$ of $S^3$ appears. We introduce
$F^{on}_{CFT}=F^{on}_{ub}+F^{on}_{sb}$ with
$F^{on}_{ub}=V_3\kappa\hat{r}^2/\rho$ and
$F^{on}_{sb}=-\hat{r}^2F^{on}_{ub}$. As an example, we have a definite relation of $
V_3\kappa=\pi l^3/8G_5=N^2/4$ for ${\cal N}=4, SU(N)$ SYM
theory\cite{Gub1,Gub2}. Then its entropy takes the form of $S_{SYM}=\pi
N^2$ at $\hat{r}=1$.

 We call these  either the UV/IR
scaling transformation in the AdS/CFT correspondence or the Tolman
red-shift transformation on the gravity side~\cite{GPP}. The
scaling factor of $\sqrt{-g^{tt}_{\infty}}=l/\rho$ comes from the
fact the Killing vector $\partial/\partial t$ is normalized so
that near infinity \beq \label{5eq7}
g\Big(\frac{\partial}{\partial t},\frac{\partial}{\partial t}\Big)
\to -\frac{\rho^2}{l^2}.\eeq
 This fixes
the  red-shifted CFT of $T_{CFT},~E_{CFT}$ and $F^{on}_{CFT}$, but
the entropy remains unchanged  under the UV/IR transformation. On
the CFT side, we find the minimum temperature
$T^0_{CFT}=\sqrt{2}/\pi \rho$ at $\hat{r}=\hat{r}_0=1/\sqrt{2}$
and the critical temperature $T^1_{CFT}=3/2\pi \rho$  from
$F^{on}_{CFT}=0$ at $\hat{r}=\hat{r}_1=1$ with $\rho=1$. These are
useful for describing  the confining/deconfining transition on the
boundary.  There exists a phase transition from bubble to a
radiation-like matter on the boundary, as contrasted with the Hawking-Page transition
in the bulk. Also we have
an approximate  relation between entropy and energy: $S_{CFT} \sim
(E_{CFT})^{3/4}$ which is the same form as in the large AdS$_5$
black hole.

 From the thermal equilibrium condition of $T=T_{CFT}$,
we find two roots of unstable bounce (UB) and stable bounce (SB):
\beq \label{twor} \hat{r}_u=\pi \rho T-\sqrt{(\pi \rho T)^2-2},~
\hat{r}_s=\pi \rho T+\sqrt{(\pi \rho T)^2-2}. \eeq For $T
\gg 1/ \pi \rho$, we have the limiting forms of $\hat{r}_u \to 0$
and $\hat{r}_s \to 2 \pi \rho T$. At $T=T^0_{CFT}$, we find an
extremum at $\hat{r}=\hat{r}_0(=\hat{r}_u=\hat{r}_s)$. For
$T>T^0_{CFT}$, string theory backgrounds of a bubble at $\hat{r}=0$,
UB and SB appear as saddle points in the Euclidean path integral of
Yang-Mills theory\cite{AGLW}. The Euclidean UB undergoes the
Gross-Witten phase transition in the large $N$ limit at a
temperature below the Hagedorn temperature. This phase transition
is identified with the Horowitz-Polchinski correspondence point
where the size  of UB becomes comparable to string scale
$(\hat{r}_u \sim l_s/l)$.

For $\hat{r}>1$ (high temperature limit of ${\cal N}=4$ SYM
theory), from Eqs.(\ref{3eq7}) and (\ref{4eq7}),  we have the
well-known form for  on-shell free energy \beq \label{hfree}
 \rho F^{on,high}_{CFT}\simeq \frac{ N^2}{4}\Big[
-\Big(\pi \rho T_{CFT}\Big)^4+ \Big( \pi \rho T_{CFT}\Big)^2\Big].
\eeq  The Gibbs free energy is defined by
$G_{CFT}=F^{on}_{CFT}+p_{CFT}V=2
 V_3 \kappa \hat{r}^2/\rho$ with the equation of state $p_{CFT}=E_{CFT}/3V$ and
 the volume of the boundary system
$V=V_3\rho^3$.  Then the  Casimir energy  is given by
$E_c=3G_{CFT}$, which  is also given by $E_c=(l/\rho)E_{BH}^c$. This shows a feature of thermal CFT defined on the
compact manifold $S^3$. Now we obtain  the Cardy-Verlinde formula
instead of $S_{CFT} \sim (E_{CFT})^{3/4}$\cite{KPSZ} \beq
\label{6eq7} S_{CFT}=\frac{2 \pi \rho}{3}\sqrt{E_c(2E_{CFT}-E_c)}
\eeq which indicates the  exact  relation between the entropy and
energy. Since $G_{CFT}= E_{c}/3 =S_c/2 \pi \rho$, we obtain
$c_{CFT}=S_c/4 \pi=  V_3\kappa \hat{r}^2$ as  a generalized
central charge at the boundary CFT.   We recover the central
charge for large $N,~{\cal N}=4$ SYM theory \beq \label{central}
c_{SYM}=\frac{N^2}{4} \eeq at the confining/deconfining transition point of
$\hat{r}=1(r_+=l)$. The exact form is given by
$c_{CFT}=(N^2-1)/4$.  Considering the confining/deconfining transition at
the critical temperature $T=T^1_{CFT}(\hat{r}=1)$, we observe the
changes of boundary quantities:
\begin{eqnarray} \label{CV1}
 &E_{CFT}&= E_c=S_{CFT}=0,~F^{on}_{CFT}>0, ~~{\rm
 for}~\hat{r}<1\\ \label{CV2}
&E_{CFT}&= E_c=T^1_{CFT}S_{CFT},~F^{on}_{CFT}=0, ~~{\rm
 for}~\hat{r}=1 \\ \label{CV3}
  &E_{CFT}& >E_c,~S_{CFT}<(1/T^1_{CFT})E_{CFT},~F^{on}_{CFT}<0, ~~{\rm
 for}~\hat{r}>1. \end{eqnarray}
  From the above, we confirm that the
Bekenstein bound of $S_{CFT} \le (1/T^1_{CFT})E_{CFT}$ is always
satisfied with the large $N$, SYM theory\cite{KPS}. Here we note
that the Cardy-Verlinde formula is suitable for only the SB. The central charge satisfies the following condition:
$c_{CFT}=0$, for $\hat{r}<1$; $c_{CFT}=c_{SYM}$, for $\hat{r}=1$;
$c_{CFT}=c_{SYM}\hat{r}^2$, for $\hat{r}>1$. Hence we argue that
for $\hat{r}\ge 1$,  the central charge is a monotonically
increasing  function of $\hat{r}$. Comparing this with the ${\cal C}_{BH}$-function on the bulk-side,
we confirm the on-shell gravity/gauge
duality.

\begin{figure}[t!]
 \centering
   \includegraphics{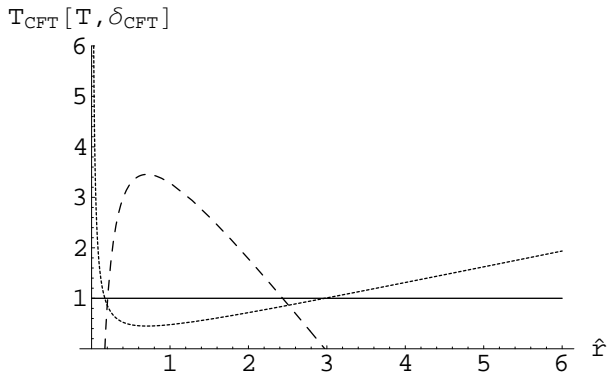} \caption{The CFT-tempearture picture of the cool (off-shell) bounce  growth
 at the Hagedorn temperature at $T>T_H$.  Solid line:
hotter temperature of heat reservoir at $T=T_s$ for Hagedorn
transition.
 Dotted line: plot of the  CFT temperature $T_{CFT}(\hat{r})=\alpha_{CFT}(\hat{r},1)$ with $\rho=1$.
 Dashed line: deficit angle $\delta_{CFT}(\hat{r},1)$.  In this case we have
 a
sequence of temperatures:
$\hat{r}_u=0.168<\hat{r}_0=0.707<\hat{r}_s=2.973$. If a matching
  occurs at $\hat{r}_{st}>\hat{r}_u$,
  the off-shell bounce  always grows into a SB.} \label{fig6}
\end{figure}

 In order to describe the Hagedorn transition on the CFT side clearly, we
need the off-shell CFT free energy
$F^{off}_{CFT}(\hat{r},T)=E_{CFT}-T  S_{CFT}$ as a
function of $\hat{r}$ and temperature $T$ of heat reservoir. Its
CFT action is given by $I_{CFT}=\beta F^{off}_{CFT}$. We find
\beq \label{foof}
F^{off}_{CFT}(\hat{r},T)=F^{off}_{ub}+F^{off}_{sb}, \eeq
where \beq \label{offcf} F^{off}_{ub}=3F^{on}_{ub}
\Big[1-\frac{2}{3}\frac{T}{T_{CFT}}\Big],~~
F^{off}_{sb}=-3F^{on}_{sb}\Big[1-\frac{4}{3}\frac{T}{T_{CFT}}\Big].
\eeq This  is identical with that constructed by using the Landau
theory of phase transition\cite{CM}.  Actually,
$F^{off}_{CFT}$ describes the CFT system away from
equilibrium ($T\not=T_{CFT}$). We confirm that $F^{on}_{CFT}$ is
the on-shell free energy by using operation: $\partial
F^{off}_{CFT}/\partial \hat{r}=0 \to T=T_{CFT} \to
F^{off}_{CFT}=F^{on}_{CFT}$. To capture the off-shell
feature, we introduce the off-shell parameter $\alpha_{CFT}$
related the deficit angle $\delta_{CFT}$
 \beq \label{off-shellc} \alpha_{CFT}(\hat{r},T) =
\frac{T_{CFT}(\hat{r})}{T}\equiv
1-\frac{\delta_{CFT}(\hat{r},T)}{2 \pi}. \eeq Here the deficit
angle takes a value between
$\delta^{min}_{CFT}=0(\alpha_{CFT}=1,T=T_{CFT})$ and
$\delta^{max}_{CFT}=2(\pi-\sqrt{2})(\hat{r}=\hat{r}_0)$. See Fig. 6. Then the
off-shell free energy takes the form \beq \label{defreec}
F^{off}_{CFT}(\hat{r},T)=\frac{F^{on}_{CFT}(\hat{r})}{\alpha_{CFT}}
+\frac{(\alpha_{CFT}-1)}{\alpha_{CFT}}E_{CFT}(\hat{r}), \eeq where
we check that for $\alpha_{CFT}=1$, we
recover $F^{off}_{CFT}=F^{on}_{CFT}$. For fixed
$T\not=T_{CFT}$, $F^{off}_{CFT}$ indicates  off-shell
configurations in the boundary CFT. Hence we expect that in the
Hagedorn transition at $T=T_s>T_1$, the off-shell free energy
$F^{off}_{CFT}$ describes a process of tachyonic
condensation.
\begin{figure}[t!]
   \centering
\includegraphics{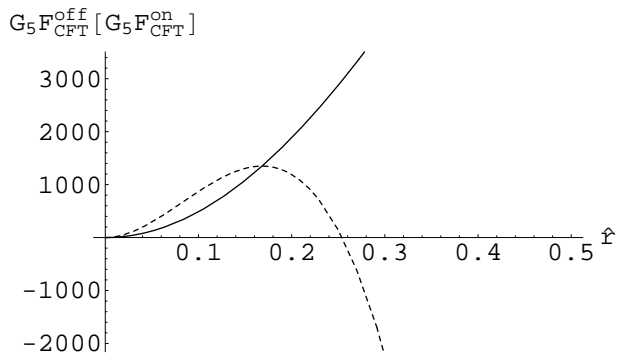}
\caption{The CFT-free energy picture of  cool (off-shell) bounce
growth in a heat reservoir at $T=T_s$ for $\hat{r}<1$.} \label{fig7}
\end{figure}

\begin{figure}[t!]
   \centering
   \includegraphics{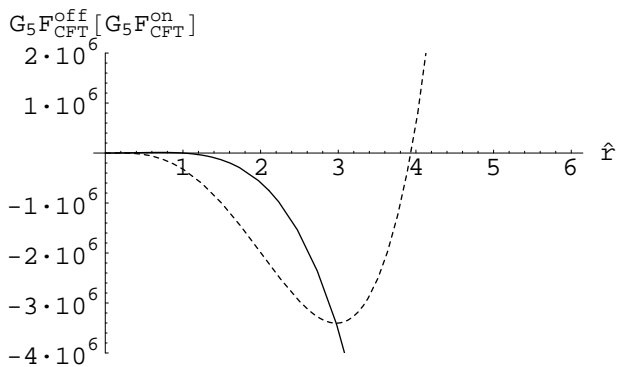}
\caption{The CFT-free energy picture of  cool (off-shell) bounce
growth in a heat reservoir at $T=T_s$. } \label{fig8}
\end{figure}

In Fig. 7, the solid line represents the on-shell CFT free energy
   $F^{on}_{CFT}(\hat{r})\simeq F^{on}_{ub}(\hat{r})$ with  $\rho=1$. The
dashed line denotes the off-shell CFT free energy
$F^{off}_{CFT}(\hat{r},1)=F^{off}_{ub}(\hat{r},1)$,
where the maximum is at $\hat{r}_u$. As is shown in Fig. 8, the solid line represents the on-shell free
energy $F^{on}_{CFT}(\hat{r})$ which is the maximum at
$\hat{r}=\hat{r}_0$ and zero at $\hat{r}=\hat{r}_1=1$. The
dashed line denotes the off-shell free energy
$F^{off}_{CFT}(\hat{r},1)\simeq F^{off}_{sb}(\hat{r},T=1)$ to
connect $\hat{r}_u$ with  $\hat{r}_s$ in the units of $G_{5}$. At
$\hat{r}=\hat{r}_s$, we have the endpoint of Hagedorn
tachyon condensation as a  SB. Comparing these with the bulk results,
we confirm the gravity/gauge duality.

\begin{figure}[t!]
 \centering
   \includegraphics{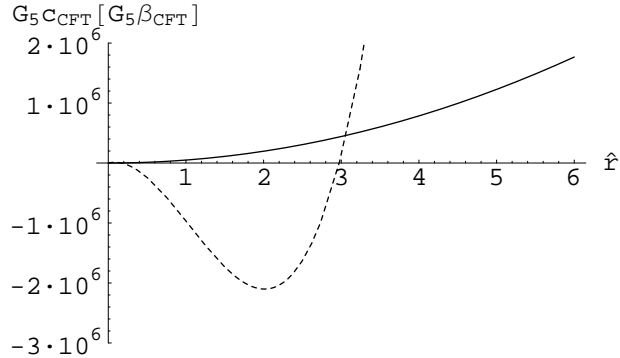} \caption{The central chrage function $c_{CFT}(\hat{r})$
   and  the CFT $\beta$-function.
The solid line represents the (on-shell) central charge
$c_{CFT}(\hat{r}\ge 1)$ which is  a monotonically
 increasing function, while the dashed line denotes the off-shell
 $\beta_{CFT}(\hat{r},1)$. The latter is zero when $\delta_{CFT}=0$. } \label{fig9}
\end{figure}

In thermodynamics, $\Delta G_{CFT} \ge 0$ holds for irreversible
processes. Similarly  $\Delta c_{CFT} \ge 0$ shows a
thermodynamic Zamolodchikov's theorem.
 As is shown Fig. 9, the central
charge function $c_{CFT}(\hat{r} \ge 1)$ is a monotonically
increasing function ($\nearrow$). The on-shell free energy
$F^{on}_{CFT}(\hat{r})$ is a monotonically decreasing function ($\searrow$).
Hence we have a monotonicity  between $c_{CFT}(\hat{r})$
with $\nearrow$ and $F^{on}(\hat{r})$ with $\searrow$.

Now we discuss the off-shell picture.  From the definition of the
CFT $\beta$-function, we have \beq \label{betafucnc}
\beta_{CFT}(\hat{r},T) \propto \frac{\partial I_{CFT}}{\partial
\hat{r} }=-6 c_{CFT}(\hat{r})\delta_{CFT}(\hat{r},T). \eeq As is
shown in Fig. 9, this CFT $\beta$-function (holographic RG-flow)
is similar to the $\beta$-function on the world sheet (RG-flow).
 Thus the holographic
RG-flow is just an irreversible thermal process in the boundary
thermal CFT.
On the other hand,
 the off-shell free energy $F^{off}_{CFT}(\hat{r},1)$
is a monotonically decreasing function, which is responsible for
the off-shell bounce growth. Hence we confirm the off-shell
connection between   $F^{off}_{CFT}(\hat{r},1)$ with $\searrow$ and $\beta_{CFT}(\hat{r},1)$ in the shape of
 $\searrow
\nearrow$.

Finally, we comment on the similarity and difference between the bulk quantities in sections 2 and 3,
and the boundary quantities in this section. First we observe a close similarity  between two systems from
Eqs. (\ref{3eq7}) and (\ref{4eq7}). It is obvious from the holographic principle that there exist
bulk-boundary relations for thermodynamic quantities. Therefore, we observe the similarity between Figs. 3, 4, 5
and Figs. 7, 8, 9. However, we find crucial differences. First of all, we could not define the black hole pressure
(presumably, $p_{BH}=0$), while the pressure of the CFT  is $p_{CFT}=\rho_{CFT}/3$ in the high temperature limit of ${\cal N}=4$ SYM theory.
That is, the CFT plays a role of the radiation-like matter. This means that we may read off
the pressure of the black hole from its dual CFT. The Cardy-Verlinde formula in Eq. (\ref{6eq7}) was originally
defined for the gauge theory of CFT on the boundary~\cite{Ver}. As an analogy,
we define the Cardy-Verlinde formula in Eq. (\ref{BCV}) on the bulk-side.
Hence the meaning of  the central charge $c_{CFT}$ is more clear than that of ${\cal C}_{BH}$-function.
The former shows the number of degrees of freedom, while the latter is just a function related to the heat capacity
of the black hole. In addition, the $F^{off}_{BH}$ and $\beta_{BH}$ describes the growth of  a black hole with the conical singularity at the horizon,
while $F^{off}_{CFT}$ and $\beta_{CFT}$ describe the growth of the bounce on the boundary.

\section{d=10 Schwarzschild black hole in a cavity}

Now we study the tachyon condensation  in the d=10 Schwarzschild
black hole (bh)\cite{BB1}. Here we achieve the IR regularization by
introducing a confining cavity instead of AdS
regularization\footnote{The AdS regulator is $l$, whereas the d=10
black hole regulator is the radius  $R$ of the cavity.}. Then we may
 connect the d=10 Hagedorn strings to a black hole
instability that ends in the large, stable black hole.

 We start with the d=10 Schwarzschild black hole spacetime
\beq \label{1eq6}
ds^2_{10Sch}=-\Big[1-\Big(\frac{r_+}{r}\Big)^7\Big]dt^2+\frac{dr^2}{1-\Big(\frac{r_+}{r}\Big)^7}
+r^2d\Omega_8^2\eeq whose thermodynamic quantities as measured at
infinity  are given by \beq \label{thermo1}
E^{\infty}=M=\frac{V_8r_+^7}{2 \pi G_{10}},~
T^{\infty}_{H}=\frac{7}{4\pi
r_+},~C^{\infty}_{bh}=-\frac{2V_8r_+^8}{G_{10}} \eeq with the
volume of the unit eight sphere $V_8=32 \pi^4/105$ and  the d=10
Newton constant $G_{10}$.
 The bh-free energy and
generalized free energy are
 \beq \label{thermo2}
F^{\infty}_{bh}=\frac{V_8r_+^7}{16 \pi
G_{10}},~F^{\infty}=E^{\infty}-T\cdot
S_{bh}=M\Big[1-\frac{7}{8}\frac{T}{T^{\infty}_H}\Big],~S_{bh}=\frac{V_8r_+^8}{4
G_{10}}. \eeq It is obvious from Eqs.(\ref{thermo1}) and
(\ref{thermo2}) that the d=10 Schwarzschild black hole is unstable
to decay into thermal gravitons  because  $C^{\infty}_{bh}<0$
and $F^{\infty}_{bh}>0$. Also we have the relation of $S_{bh} \sim
(E^{\infty})^{8/7}$, which signals the presence of an unstable
canonical ensemble. Considering $E_c^{\infty}=9E^{\infty}-8T_H^{\infty}\cdot S_{bh}=2E^{\infty}$\cite{KPSZ},
we have the exact relation between entropy and energy: $S_{bh}=\frac{2\pi r_+}{8}\sqrt{2E^{\infty}E_c^{\infty}}$.
Comparing this with Eq.(\ref{BCV}) leads to the conclusion that the Schwarzschild black hole is unstable.
 This black hole can be rendered
thermodynamically stable by confining it within a finite ideal
isothermal cavity. We assume that a black hole is located at the
center of the cavity. Here we fix the temperature $T$ on its
isothermal boundary of radius $R$. In an  equilibrium configuration,
the Hawking temperature measured on the boundary must be equal to
the boundary temperature $T$\cite{York} \beq T_{H}(R,r_+)\equiv
\frac{T_{H}^{\infty}}{\sqrt{1-\Big(\frac{r_+}{R}\Big)^7}}=T.
\label{2eq6}
 \eeq
 \begin{figure}[t!]
 \centering
   \includegraphics{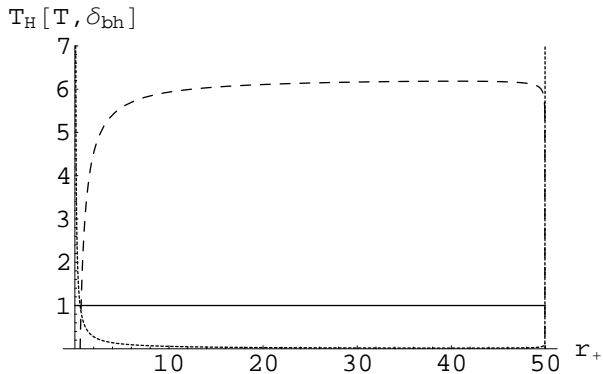} \caption{The tempearture picture of the d=10 cool (off-shell) black hole growth
 in a cavity at the Hagedorn temperature at $T>T_H$. Dashed
line: the deficit angle $\delta_{bh}(R,r_+,1)$. Solid line: hotter
temperature of heat reservoir at $T=T_s$ for Hagedorn
transition.
 Dotted line:  the  Hawking temperature $T_H$ of the  cold off-shell black hole. In this case we have
 a sequence of temperatures: $r_u=0.557<r_0=40.33<r_s=49.99$.  If a
matching
  occurs at $r_{st}>r_u$, the stringy black hole always grows into a large, stable black
hole at $r_+=r_s$.} \label{fig10}
\end{figure}
This means that, according to  the Tolman law,  a local observer
at rest will measure a local temperature $T$ which scales as
$1/\sqrt{-g_{00}}$ for any self-gravitating system in thermal
equilibrium with heat reservoir. The cavity is  the
heat reservoir. At $r_+=r_0=(2/9)^{1/7}R$, $T_H$ has the minimum
temperature
 \beq
 \label{3eq6}
 \tilde{T}_0=\frac{3\sqrt{7}}{4\pi R}\Big(\frac{9}{2}\Big)^{\frac{1}{7}}
 \eeq
 which corresponds to the nucleation temperature of a stable black hole.
This is depicted in Fig. 10. The  equation (\ref{2eq6}) allows two real, nonzero
solutions for a given $T$: a smaller unstable black hole (ubh)
with radius  $r_u$ and a larger, stable black hole (sbh) with  radius
$r_s$. For $T<T_0$, no real value $r_+$ can solve
Eq.(\ref{2eq6}) and thus no black hole can exist in the
cavity. The thermodynamic energy for the black hole embedded in a
cavity takes the form of
$E_{bh}(R,r_+)=E^{max}\Big(1-\sqrt{1-r_+^7/R^7}\Big)$ which
differers from the ADM mass of $E^{\infty}=M$. Here
$E^{max}=V_8R^7/\pi G_{10}$ is the maximum energy of this system
when $r_+=R$ (system-size black hole). In general, one has $0 \le
E \le E^{max}$. The heat capacity is defined at the constant area
$A$ of the cavity boundary. It is given by $C_{bh}\equiv (\partial
E_{bh}/\partial T_{H})_A$\cite{York} \beq
 \label{4eq6}
C_{bh}(R,r_+)=C^{\infty}_{bh}\frac{\Big[1-\Big(\frac{r_+}{R}\Big)^7\Big]}
{\Big[1-\frac{9}{2}\Big(\frac{r_+}{R}\Big)^7\Big]}.
 \eeq
 The bh-free energy of the system is given by $F^{on}_{bh}=E_{bh}-T_{H}S_{bh}$.
 $F_{bh}=0$
leads to $r_+=r_1=(32/81)^{1/7}R$. Substituting it into
Eq.(\ref{2eq6}), the transition temperature can be obtained as
\beq \label{6eq6} \tilde{T}_1=\frac{1}{2 \pi
R}\Big[\Big(\frac{9}{2}\Big)^{\frac{1}{7}}+2\Big(\frac{2}{9}\Big)^{\frac{1}{7}}\Big].
\eeq $C_{bh}$ has an unbounded discontinuity at $r_+=r_0=40.33$
for $R=50$.
 Therefore, it seems that the assumed phase transition
is first-order at $T=\tilde{T}_0=0.0156$. However, this behavior
of heat capacity in itself does not indicate a phase transition in
the canonical ensemble. The heat capacity usually determines
thermal stability of the system. Here we find the positive heat
capacity for $r_0 \le r_+ \le R$, indicating stability.   On the
other hand, the bh-free energy $F^{on}_{bh}$ has the maximum value
at $r_+=r_0$ and it becomes zero at $r_+=r_1=43.78$. The latter
provides $\tilde{T}_1=0.0163$ which is the critical temperature
for the GPY phase transition.

\begin{figure}[t!]
\centering
   \includegraphics{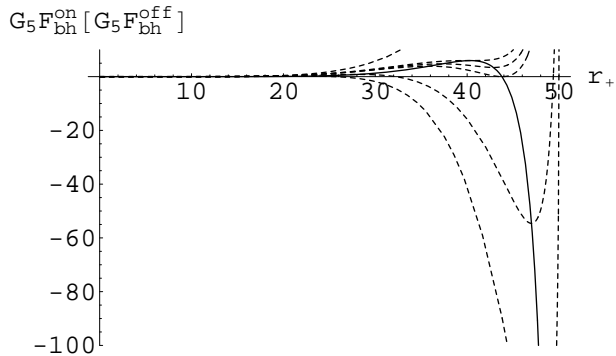}
   \caption{The process of   black hole nucleation as the GPY phase transition.
  The the solid line represents the bh-free
energy $F^{on}_{bh}(R=50,r_+)$, while the dashed lines denote the
generalized free energy $F^{off}_{bh}(R=50,r_+,T)$ in the units of $G_{5}$.
At $r_+=r_0$, $F^{on}_{bh}$ has the maximum value and it is
zero at $r_+=r_1=43.78$.
 From the top down, we have the generalized free energy graphs
for $T=0.01, \tilde{T}_0(=0.0156), 0.016, \tilde{T}_1(=0.0163),
0.02, 0.025$.} \label{fig11}
\end{figure}

To study the GPY and Hagedorn  transitions explicitly, we need to
introduce the generalized free energy\cite{York,SH}  \beq
\label{7eq6} F^{off}_{bh}(R,r_+,T)=E_{bh}-T S_{bh}. \eeq The
generalized free energy $F^{off}_{bh}$ plays a role of an
effective potential in the canonical ensemble.  Here we define the
action $I_{bh}=\beta F^{off}_{bh}$.   The
d=10 deficit angle is defined by (see Fig. 10) \beq \label{d=10def}
\delta_{bh}(R,r_+,T)=2 \pi\Big[1-\frac{T_H(R,r_+)}{T}\Big]. \eeq
With $T=\tilde{T}_0$, an
extremum appears at $r_+=r_0(=r_u=r_s)$. It could be checked by
noticing an inflection point in Fig. 11. For $T>\tilde{T}_0$, there
are two extrema, the ubh with radius $r_u$ and the sbh with $r_s$.
We note that for $\tilde{T}_0<T<\tilde{T}_1$, $F^{off}_{bh}$ has a
saddle point at $r=r_u$. This unstable solution is important as
the mediator of phase transition from thermal gravitons to a sbh.
In the limit of $R\to \infty$, the sbh is lost and only the usb
survives. $F^{on}_{bh}$ is a set of saddle points of
$F^{off}_{bh}$. Now we examine how the tachyon condensation could
be  realized in this picture\cite{HS}. There exists a slight difference
in including winding number between AdS and cavity regularization.
The Euclidean topology of d=10 Schwarzschild black hole in cavity
is $R^2 \times S^8$, which is not the cylindrical topology of
$S^1 \times R^9$. Hence the  d=10 Schwarzschild black hole in
cavity seems to be unable to include the Hagedorn  tachyon
condensation, because it does not have a non-contractible circle.
Inside the horizon, cylinders with spherical cross sections shrink.
Hence it seems that a topologically stable winding tachyon does not
appear. Despite this, there may exist a superposition of winding
string modes in great circles of  $S^8$. $S^8$ starts shrinking
rapidly before its spatial curvature becomes large. The change in
radius $r$ with respect to proper time is given by $\dot{r}=-
\sqrt{(r_+/r)^7-1}$, which takes the form of $-(r_+/r)^{7/2}$ for
$r \ll r_+$. This means that the velocity $\dot{r}$ becomes very
rapid for $l_s \ll r \ll r_+$, while $S^8$ is still large.
Starting from the radius $r_c$ such that
$(\dot{r}/r)^2|_{r=r_c}=1/l_s^2$, one finds a capping radius
$r_c=(r_+^7l_s^2)^{1/9}$. This implies that the Hagedorn density
of strings is produced by the time the sphere has shrunk to
$r=r_c$.  The backreaction of this string gas may behave like a
winding tachyon condensation\cite{BPR1,BPR2}. When the size of horizon
reaches the string scale ($r_+ \sim l_s$), it will pinch off at
$r_c=l_s$, removing the region of  large curvature sector from
the small sector including the  singularity. This  is a process of
tachyon condensation in d=10 Schwarzschild black hole. Considering
the d=10 Schwarzschild black hole in a cavity, the large sector
grows into a sbh (cavity-size black hole).

We discuss the Hagedorn transition by introducing an approximate
off-shell free energy to Eq.(\ref{7eq6}), \beq \label{appfe}
F^{off}_{app}=8F^{on}_{ubh} \Big[ \Big(1-\frac{7}{8}\frac{T}{T_{H}}\Big)
+\frac{1}{4}\Big(\frac{r_+}{R}\Big)^7
\Big(1-\frac{7}{4}\frac{T}{T_{H}}\Big)\Big],~~F^{on}_{ubh}=\frac{V_8r_+^7}{2
\pi G_{10}}=M \eeq which holds for $r_+ <R$. In the case of $r_+\ll
R$, this leads to  the off-shell free energy $F^{\infty}$ for the
d=10 Schwarzschild back hole. In the case of $r_+ \sim R$, the
above free energy does not work for the Hagedorn transition
because we have to include all higher-order terms of $(r_+/R)^{7n},~n \ge 2$. However, we  find a
similarity between Eq.(\ref{8eq4})  and Eq.(\ref{appfe}).

 In order to estimate where the ubh and sbh are located, we use
Fig. 12 and Fig. 13. From
$F^{off}_{BH}(R=50,r_+,1)=F^{on}_{bh}(R=50,r_+)$, we obtain
the ubh
 with size $r_u$ and sbh with size $r_s\simeq R$. The latter is a nearly
cavity-size black hole. As is shown in Fig. 12,  we observe the
barrier of the ubh at $r_u$, even though its maximum is very
small as $G_{5}F^{\infty}(0.557,1)=1.2 \times 10^{-12}$. Also we
observe the
 sensitivity around $r_+=l_s=1$ that
$G_5F^{\infty}(R^J_{st}=0.318,1)=8.9 \times
10^{-14},~G_5F^{\infty}(0.636,1)=-13.4 \times
0^{-14},~G_5F^{\infty}(1,1)=-2.7 \times 10^{-10},$ and $
G_5F^{\infty}(2,1)=-1.3 \times 10^{-7}$. The distance $\Delta
r_+=0.079$ between the maximum and zero is the same  as the AdS
black hole. Also we observe that the off-shell free
energy is very sensitive around the string scale $r_+=l_s$.

\begin{figure}[t!]
   \centering
   \includegraphics{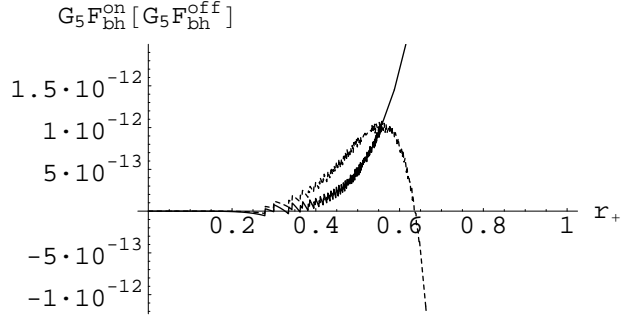}
\caption{Hagedorn transition with $T=T_s$ for $r_+<1$.
   The solid line represents the bh-free
energy $F^{on}_{bh}(R=50,r_+)\simeq F^{\infty}_{bh}(r_+)\sim {\cal C}_{bh}(R=50,r_+)$, while
the dashed line denotes the generalized free energy
$F^{off}_{bh}(50,r_+,1)\simeq F^{\infty}(r_+,1)$. This picture
is  described by a small Schwarzschild black hole.}
\label{fig12}
\end{figure}

\begin{figure}[t!]
   \centering
   \includegraphics{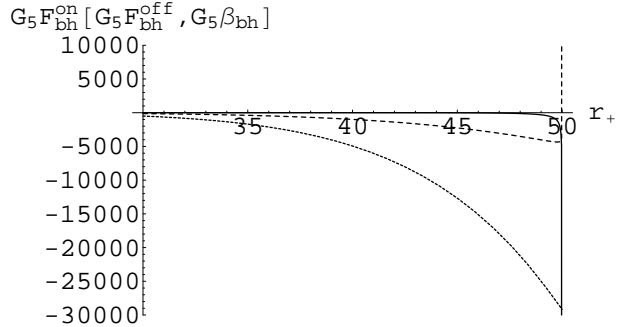}
\caption{The free energy picture of the  d=10 off-shell black hole
growth
  in a cavity at $T=T_s$ for $r_+ \gg 1$.
   The solid line represents the on-shell free
energy $F^{on}_{bh}(R=50,r_+)$, while the dotted line denotes the
off-shell free energy $F^{off}_{bh}(50,r_+,1)$ in the units of
$G_{5}$. The dashed line represents the off-shell $\beta$-function
$\beta_{bh}(50,r_+,1)$. At $r_+=r_s=49.99$, we have the
endpoint of Hagedorn tachyon condensation as a globally stable
black hole state.  } \label{fig13}
\end{figure}

We assume that the cavity is filled with thermal gravitons.
Considering $S_{TG}\sim (TR)^9$ and $RE_{TG} \sim (TR)^{10}$, one
finds the relation of $S_{TG} \sim (E_{TG})^{9/10}$.  We obtain
the collapsing temperature $T=\tilde{T}_2$ approximately by using
the relation \beq \label{8eq6} E_{TG}= \tilde{T}_2^{10}R^9 \sim
E_{bh}\simeq R^7/G_{10} \eeq which shows that $\tilde{T}_2 \sim
1/(G_{10}R^2)^{1/10}$.
 At $T=\tilde{T}_2=0.148$, the
gravitational collapse of thermal graviton may occur.  The
collapsing temperature is  given by  $\tilde{T}_2 \sim 10
\tilde{T}_1$ and the Hagedorn temperature is  given by $T_s
\sim 100\tilde{T}_1$.
 Hence we are not easy to avoid  this collapsing  from our
consideration, in contrast with the thermal AdS collapse at
$T_2=2.195>T_s$ because of
$\tilde{T}_2\in(\tilde{T}_1,T_s)$. It seems that the
classical gravitational collapse  occurs by the relativistic Jeans
instability before the Hagedorn transition at $T=T_s$.

Finally, we discuss the connection between the on-shell and
off-shell quantities. Since the dual cft to the asymptotically
flat d=10 black hole is not yet known, we cannot precisely define
the Casimir energy $E_{bh}^{c}$ and thus  ${\cal C}_{bh}$. However, assuming $E^{c}_{bh}=2E^{\infty}$, we
may define ${\cal C}_{bh}(R,r_+)=E_{bh}^{c}/4\pi \tilde{T}_1 \sim
V_8Rr_+^7/\pi G_{10}$, which is a monotonically increasing
function.  Now let us derive
the corresponding $\beta$-function \beq \label{d=10beta}
\beta_{bh}(r_+,T) \propto \frac{\partial I_{bh}}{\partial r_+}
\sim -\frac{{\cal C}_{bh}(R,r_+)}{R} \delta_{bh}(r_+,T). \eeq
Hence we have a monotonic connection between ${\cal
C}_{bh}(R,r_+)$ with $\nearrow$ and $F^{on}_{bh}(R,r_+)$ with
$\searrow$ shown in Fig. 13. Further we confirm the off-shell
connection between $F^{off}_{BH}(R,r_+,1)$ with $\searrow$ and  $\beta_{bh}(R,r_+,1)$ in the shape of  $ \searrow
\uparrow$.

\section{Summary}
We study  quasilocal tachyon condensation by using the
gravity/gauge duality.
In order to cure the IR divergence due to
tachyon, we introduce two regularization schemes: AdS space and
Schwarzschild black hole in a cavity, which  provide stable
canonical ensembles and thus are good candidates for the endpoint of
tachyon condensation. Introducing the Cardy-Verlinde formula, we
establish the on-shell gravity/gauge duality. The Cardy-Verlinde
formula states exactly the on-shell relationship between the
entropy and energy instead of an approximate relation $S \sim
E^{3/4}$ for the AdS black hole and its CFT.
\begin{table}
 \caption{Summary for  the Hagedorn  transitions at $T=T_s$ in the  study of tachyon condensation.
 Here is the notation: SP(saddle point), UBH(AdS unstable black hole), SBH(AdS stable black hole), UB(unstable bounce),
 SB(stable bounce), sbh(d=10 stable black hole), ubh(d=10 unstable black hole), and WST(world sheet topology). }

\begin{tabular}{|c|c|c|c|c|}
  \hline
   & string theory & AdS-BH  & CFT & d=10 bh  \\
  \hline
  on-shell instability & SP(hole) & UBH & UB & ubh  \\ \hline
  on-shell flow  & N/A &${\cal C}_{BH}/F^{on}_{BH}$& $c_{CFT}/F^{on}_{CFT}$ &  ${\cal C}_{bh}/F^{on}_{bh}$ \\ \hline
  off-shell matching & tachyon & BH & bounce & bh  \\ \hline
  off-shell flow  & $\beta/V_{eff}^t$ & $\beta_{BH}/F^{off}_{BH}$ & $\beta_{CFT}/F^{off}_{CFT}$
   & $\beta_{bh}/F^{off}_{bh}$ \\ \hline
  off-shell feature& WST change & BH growth & bounce growth & bh growth \\ \hline
  on-shell stability & SP(hole) & SBH & SB & sbh \\ \hline

\end{tabular}
\end{table}
We summarize our key results in Table 1.
For the tachyon condensation, the on-shell flow is not available. In string theory,
the RG $\beta$-function shows a collection of off-shell configurations on the world sheet.
Further, an effective tachyon potential is given by Eq.(\ref{7eq7}).
First, we note the on-shell correspondence: ${\cal C}_{BH} \to c_{CFT} \to {\cal C}_{bh}$
for the monotonic increasing  $C$-function ($\nearrow$);
$F^{on}_{BH} \to F^{on}_{CFT} \to F^{on}_{bh}$ for the monotonic decreasing  free energy $F^{on}$ ($\searrow$).
Then the off-shell correspondence is as follows: $F^{off}_{BH} \to F^{off}_{CFT} \to F^{off}_{bh}$
 for the monotonic decreasing  free energy $F^{off}$ ($\searrow$) ; $\beta_{BH} \to \beta_{CFT} \to \beta_{bh}$
for $\beta$-function ($\searrow \nearrow$).
The connection between these can be found as well.
The $\beta$-function is  the derivative of $F^{off}$ with respect to $r_+(\hat{r})$ and measures
the mass $M_{cs}\simeq \beta/16\pi$ of the conical singularity at the event horizon.
The on-shell free energy $F^{on}$ is obtained by substitution of $T\to T_H^A(T_{CFT},T_H)$ into $F^{off}$.
Also this can be seen  from the derivative of $F^{off}$
with respect to $r_+(\hat{r})$.
The $C$-function is proportional to the unstable part $F^{on}_{UBH}(F^{on}_{ub},F^{on}_{ubh})$.

A few of comments are in order.
 The
stringy geometry resulting from the off-shell tachyon dynamics
 matches onto the off-shell AdS black
hole at the matching radius $r_+=r_{st}\in \{l_s,l\}$. In choosing
this region, we avoid including  the Jeans instability
 at $r_{st}^J=0.318$, but we take into account  the
Horowitz-Polchinski correspondence point for black holes and
strings at $r_{HP} \sim l_s$ and the tachyon instability due to
the winding number at string scale $r_+ \sim l_s$.  If the matching occurs at
$r_+=r_0=l/\sqrt{2}$, we find  the maximal deficit angle
$\delta(X_0)\simeq 2\pi$. The amount of free energy
released at the moment of the matching is $F^{off}_{BH}(50,r_+=r_0,1)=-215881/G_5 \simeq
-10N^2$.  Furthermore, the matching
point of $r_{st}=l_s$ could be extended to $r_{st}={\cal O}(l)$
unless $\beta_s-\beta$ is fine-tuned to be very small\cite{BB1}.
For $r_{st}=l_s$, the released free energy  is given by
$F^{off}_{BH}(50,r_+=1,1)=-35/G_5\simeq 10^6$, whereas for $r_{st}=l$, the
released free energy is
$F^{off}_{BH}(50,r_+=l,1)=-610960/G_5\simeq 30N^2$.
 Hence, we confirm that for $r_{st} \propto {\cal O}(l)$,
the released free energy is of order  $N^2$\cite{BB21,BB22}.

 Concerning the weakly coupled case, it is known that
the phase transition is similar to  the strongly coupled case. Of
course, the would-be black hole is different from the AdS black
hole for the strongly coupled case\cite{AMMPR}.  Fortunately, the
corresponding Cardy-Verlinde formula takes the same form as
Eq.(\ref{6eq7}) replacing $2 \pi\rho/3$ and $\hat{r}=r_+/l$
by $4 \pi\rho/3$ and $\delta^{-1}=2 \pi \rho T$ from the free
${\cal N}=4$ SYM theory\cite{KPSZ}. This shows that ``2" for $\lambda^{1/4}
\to \infty$ (in the strong coupling limit), while ``4" for
$\lambda^{1/4} \to 0$ (in the weak coupling limit).

Further, the d=10 Schwarzschild black hole in a cavity is
considered a model for the Hagedorn transition  which gives  a possible
explanation of  the tachyon condensation. However, it is not clear
 whether this is truly a model of the tachyon condensation because of the
lack of its dual cft.

In conclusion we explain the process of tachyon condensation by using
the off-shell gravity/gauge duality.

\section*{Acknowledgement}
The author thanks Steve Hsu and Brian Murray for helpful discussions.
This work was supported by the Korea Research Foundation Grant
(KRF-2005-013-C00018) and by
the Korea Science and Engineering Foundation through the Center for
Quantum Spacetime of Sogang University with grant number
R11-2005-021.

\end{document}